\documentclass[11pt]{article}
\pdfoutput=1
\usepackage{jheppub}
 \usepackage{bm}
 \usepackage{amsmath}
\usepackage{graphicx}
\usepackage{tabularx}
\usepackage{float}
\usepackage{graphicx}
\usepackage{subfig}
\usepackage{multirow}
\usepackage{makecell}
\usepackage{cleveref}
\usepackage{color, xcolor}

\usepackage{amsmath}
\usepackage{braket}
\usepackage{soul}
\usepackage{slashed}
\usepackage{url}
\usepackage{comment}
\usepackage{hyperref}

\newcommand{\ps}[1]{\textcolor{teal}{(PS: #1)}}

\newcommand{\jf}[1]{\textcolor{purple}{ #1}}

\begin{document}
\title{Probing Ultralight Higher-spin Dark Matter using Stellar Kinematics of Dwarf Galaxies}
\newcommand{\norm}{$\mu$}
\newcommand{\rcnorm}{$\psi$}
\newcommand{\trans}{$\lambda$}
\author[1]{Praniti Singh,}
\author[1,2]{JiJi Fan,}
\author[1,2]{Savvas M. Koushiappas,}
\author[3]{and Isabelle S.  Goldstein}
\affiliation[1]{Department of Physics, Brown University, Providence, RI 02912, USA}
\affiliation[2]{Brown Center for Theoretical Physics \&  Innovation, Brown University, Providence, RI 02912, USA}
\affiliation[3]{Department of Chemistry and Physics,  Monmouth University, 
West Long Branch, NJ
07764, USA}

\emailAdd{praniti\_singh@brown.edu}
\emailAdd{jiji\_fan@brown.edu}
\emailAdd{koushiappas@brown.edu}
\emailAdd{igoldste@monmouth.edu}

\abstract{
Stellar kinematics of the classical Milky Way dwarf galaxies provide a powerful probe
of the dark matter distribution and its underlying physics. In this article we use them
to explore ultralight dark matter (ULDM) with a non-zero spin. Numerical simulations in
the non-relativistic limit have shown that, for the same initial conditions, higher-spin
ULDM forms a more diffuse soliton core at the galactic center, with a smoother
transition to the NFW outer region, than scalar ULDM. We introduce a generalized
parametrization of the ULDM halo profile that captures these spin-induced effects
through three additional parameters: the normalization of the central core density, the
exponent of the core--halo mass relation, and the radius of the transition to the NFW
envelope. Applying this profile in a Bayesian analysis of six classical dwarf galaxies,
we find halo masses in the range $10^{8.5}$--$10^{9.5}\,M_\odot$ and, from each galaxy
individually, a particle mass $m \gtrsim 3\times10^{-21}$~eV. The core--halo exponent
and the transition radius are consistent with their scalar values. The core-density
normalization is also consistent with the scalar value in four galaxies and prefers a core
denser than a scalar soliton in Fornax, while the factor-of-several suppression expected
for higher-spin cores is not excluded by any single galaxy. The data therefore show no
preference for the more diffuse cores of higher-spin ULDM, but do not yet rule them out.}

\maketitle

\section{Introduction}
\label{sec:intro}

The identity of dark matter (DM) is one of the deepest mysteries in nature. Given
our limited knowledge of it, there exists a vast landscape of DM scenarios with
dramatically different predictions. One representative class of models is
ultralight dark matter (ULDM), whose theoretical motivation originated with the QCD axion.
Originally introduced to resolve the strong CP problem~\cite{Peccei:1977hh,
Weinberg:1977ma, Wilczek:1977pj}, the axion is a pseudo-Nambu-Goldstone boson
whose mass is protected by an approximate shift symmetry, and is therefore
naturally light without fine-tuning. Such a field is produced in the
early universe through the misalignment mechanism~\cite{Preskill:1982cy,
Abbott:1982af, Dine:1982ah} as a cold, spatially coherent condensate, making it a
viable DM candidate despite its tiny mass. The same logic extends far beyond the
QCD axion: axion-like particles arise ubiquitously from high-dimensional
constructions in string theory~\cite{Svrcek:2006yi, Arvanitaki:2009fg,
Cicoli:2021gss},  whose masses may span tens of orders
of magnitude and making the ultralight regime a generic, rather than exotic,
expectation. 

When dark matter is bosonic and light with mass much below $10$~eV,
the occupation number within a de Broglie volume is enormous in astrophysical and
cosmological environments.\footnote{There is not a clean  boundary between an
ultralight and light bosonic DM. For ULDM, we usually refer to bosonic DM with
mass close to the end of $10^{-22}$~eV (with an associated de Broglie wavelength
$\sim$ kpc).} Consequently, it could be described by classical waves~\cite{Hu:2000ke,
Hui:2016ltb}, a description  distinguished from the discrete point-like
particles in the cold dark matter (CDM) scenario.\footnote{For recent discussions
on the quantum effects of ULDM, see~\cite{Cheong:2024ose, Bao:2025nsd}.} ULDM,
also referred to as ``wave DM'' for reasons mentioned briefly above,\footnote{There
are quite a few names for this scenario in the literature. In addition to ULDM and
wave DM, some other names include fuzzy DM and Bose-Einstein DM.} could exhibit a
wide range of novel phenomena at various astrophysical scales, including but not
limited to suppression of small-scale structures~\cite{Matos:2000ss, Hu:2000ke}, as
well as formation of solitons~\cite{Schive:2014dra, Amin:2019ums},
turbulence~\cite{Mocz:2017wlg} and vortices~\cite{Hui:2020hbq} (in the context of
scalar ULDM). We do not intend to give a thorough review here and will refer the
reader to~\cite{Marsh:2015wka, Niemeyer:2019aqm, Ferreira:2020fam, Hui:2021tkt,
Eberhardt:2025caq} for recent reviews and references therein.

The same string theory compactifications that generate ultralight axion-like particles, however, do not produce
scalars alone: higher-dimensional gauge fields and the metric itself give rise 
to light vector and tensor states~\cite{Goodsell:2009xc, Jaeckel:2010ni}, so
the spin-0 case singled out above is only the simplest member of a broader
ultralight sector. One natural question about ULDM is therefore its underlying
spin. Indeed, beyond the extensively-studied scalar case, there exist different
production mechanisms to populate ULDM with non-zero spins, in particular vector
ULDM~\cite{Graham:2015rva, Agrawal:2018vin, Co:2018lka, Dror:2018pdh,
Bastero-Gil:2018uel, Co:2021rhi, Sato:2022jya, Adshead:2023qiw, Zhang:2025pgb}.

Fields of nonzero spin carry internal polarization degrees of freedom, and the
resulting dynamics do  not need to mirror the scalar case, so it is interesting to explore
the effects of ULDM spins at nonlinear astrophysical scales
(e.g.~\cite{Adshead:2021kvl,  Blinov:2021axd,Jain:2021pnk, Zhang:2021xxa, Gorghetto:2022sue, Amin:2022pzv, East:2022rsi,Amin:2023imi,Jain:2023ojg, Chen:2023bqy, Zhang:2023fhs, Wang:2023tly,
Chen:2024vgh, Cyncynates:2024yxm, Lopez-Sanchez:2025osk, Schiappacasse:2025mao,Amaral:2025fcd, Berger:2025pfv, Alexander:2026fam, 
Schiappacasse:2026ems, Chen:2026qos, Chen:2026dst}), with the hope of distinguishing between
the scalar and higher-spin cases. One intriguing feature, shown by numerical
simulations in the nonrelativistic limit, is that higher-spin ULDM also forms
solitonic cores as in the scalar case. Yet the quantitative details could be
different: the cores from higher-spin ULDM tend to extend further with a lower
central density, given the same initial conditions in the
simulations~\cite{Amin:2022pzv, Lopez-Sanchez:2025osk}.

One type of system which could potentially probe the variety of DM profiles from
ULDM with different spins is the Milky Way dwarf galaxies. These DM-dominated
systems, with mass-to-light ratios larger than $10$, are less subject to most
baryonic astrophysical complexities and provide ideal test systems for DM
properties. More specifically, measurements of stellar velocities along the line of
sight of bright red giants could be used to constrain the DM distribution
(e.g.,~\cite{Strigari:2006rd, Strigari:2007at, Martinez:2009jh,
2011PhRvL.107x1303G, Oswalt2013Vol5}). These measurements are powerful: under a set
of reasonable dynamical assumptions, they could be used to reconstruct the
gravitational potential robustly with well-controlled errors, especially in
classical dwarf galaxies. Indeed, this has already been applied to scalar ULDM and
sets one of the leading constraints on it~\cite{Goldstein:2022pxu,
Zimmermann:2024xvd, Teodori:2025rul, Wardana:2026qag}.

In this paper, we investigate the implications of stellar kinematics data from
Milky Way dwarf galaxies for higher-spin ULDM. We introduce a generalized
parametrization for the ULDM halo profile with additional parameters to capture the
effects induced by non-zero spins. We then implement a Bayesian analysis, explore
the allowed parameter space, and discuss its implications.

The paper is organized as follows. In Sec.~\ref{sec: stellar}, we review the basic
formalism to connect DM density profiles to the observed stellar velocity
dispersion profiles. In Sec.~\ref{sec:profile}, we discuss the generalized
parametrization of the ULDM halo profile, including the central core and outskirt
envelope. We discuss the data and present our results from the Bayesian analysis in
Sec.~\ref{sec:result}. We conclude and discuss potential future directions in
Sec.~\ref{sec:con}.

\section{Stellar Kinematics}
\label{sec: stellar}
In this section,  we review how the gravitational potential of a dwarf galaxy is constructed from its stellar velocity dispersion profile (that is, the stellar velocity dispersion as a function of projected radius from the center of the galaxy), closely following Ref.~\cite{Goldstein:2022pxu}. The stellar velocity dispersion profile is built from line-of-sight stellar velocity measurements. Readers already familiar with this formalism may safely skip ahead.

Assuming spherical symmetry, the gravitational potential $\Phi(r)$ of a dwarf galaxy can be inferred from its stellar velocity-dispersion profile using the spherical Jeans equation:
\begin{equation}
\begin{aligned}
\frac{d \Phi}{d r} & =-\frac{G M(r)}{r^2} \\
& =-\frac{1}{\nu(r)} \frac{d}{d r}\left[\nu(r) \overline{u_r^2}(r)\right]-2 \frac{\beta_a(r) \overline{u_r^2}(r)}{r}~.
\end{aligned}
\label{eqn: jeans}
\end{equation}
where $M(r)$ is the mass enclosed within a radius $r$ from the galactic center, $\nu(r)$ is the stellar density profile, and $\overline{u_r^2}(r)$ is the radial velocity dispersion of the stars. The orbital anisotropy parameter $\beta_a(r)$ measures the difference between the radial and tangential velocity dispersions:
\begin{equation}
\beta_a(r) \equiv 1-\frac{2 \overline{u_\theta^2}(r)}{\overline{u_r^2}(r)}~,
\label{eq:beta}
\end{equation}
where $\overline{u_\theta^2}(r)$ is the azimuthal velocity dispersion. The total enclosed mass generating the potential is
\begin{equation}
M(r)=4 \pi \int_0^r s^2 \rho(s) d s~,
\end{equation}
where $\rho(r)$ is the total mass density. 

We emphasize that the Jeans equation implicitly assumes dynamical equilibrium, an assumption that may not hold exactly for the Milky Way dwarf satellites considered in our analysis. Many studies have shown, however, that departures from equilibrium are unlikely to change the inferred dynamical mass appreciably \cite{10.1111/j.1365-2966.2005.09959.x,1995AJ....109.1071P,1995ApJ...442..142O}. Moreover, since the dynamical crossing time of a typical Milky Way dwarf spheroidal is much shorter than its orbital period, the system is expected to remain close to equilibrium over most of its orbit.

Under this assumption, and again assuming spherical symmetry, \cref{eqn: jeans} has the general solution 
\begin{equation}
\nu(r) \overline{u_r^2}(r)=\frac{1}{f(r)} \int_r^{\infty} f(s) \nu(s) \frac{G M(s)}{s^2} d s~,
\end{equation}
with
\begin{equation}
f(r)=2 f\left(r_1\right) \exp \left[\int_{r_1}^r \beta_a(s) s^{-1} d s\right]~,
\end{equation}
where $r_1$ is a reference scale at which the value of $f(r)$ is known.

We further assume that the orbital anisotropy is constant within a given Milky Way satellite. The mass profile $M(r)$ is related to the projected stellar density $\Sigma(R)$ and the line-of-sight velocity dispersion $\sigma^2(R)$ through projection along the line of sight:
\begin{equation}
\sigma^2(R) \Sigma(R)=2 \int_R^{\infty}\left(1-\beta_a(r) \frac{R^2}{r^2}\right) \frac{\nu(r) \overline{u_r^2}(r) r}{\sqrt{r^2-R^2}} d r~,
\label{eqn: los disp}
\end{equation}
where $R$ is the projected distance from the center of the satellite. For the stellar density profile we adopt a Plummer profile \cite{10.1093/mnras/71.5.460},
\begin{equation}
\nu(R)=\frac{3 L}{4 \pi R_e^3} \frac{1}{\left(1+R^2 / R_e^2\right)^{5 / 2}}~,
\end{equation}
which corresponds to a projected stellar density of the form
\begin{equation}
\Sigma(R)=\frac{L}{\pi R_e^2} \frac{1}{\left(1+R^2 / R_e^2\right)^2}~,
\end{equation}
where $L$ is the total luminosity and $R_e$ is the projected half-light radius.\footnote{For Plummer profile, the projected half-light radius is also called Plummer radius. We will use them interchangeably below.}

We carry out our analysis using Bayesian inference, adopting the unbinned Gaussian likelihood function of Ref.~\cite{Geringer-Sameth:2014yza},
\begin{equation}
\mathcal{L}=\prod_{i=1}^N \frac{1}{\sqrt{2 \pi} \sqrt{\delta_{u, i}^2+\sigma^2\left(R_i\right)}} \exp \left[-\frac{1}{2} \frac{\left(u_i-\langle u\rangle\right)^2}{\delta_{u, i}^2+\sigma^2\left(R_i\right)}\right],
\end{equation}
where $u_i$ and $R_i$ are the line-of-sight velocity and projected position of the $i$th star in the kinematic data, and $\delta_{u, i}$ is the corresponding observational velocity error. The bulk velocity $\langle u\rangle$ is treated as a nuisance parameter and marginalized over with a flat prior.

\section{Halo Profile}
\label{sec:profile}

To test the viability of ULDM using stellar kinematics in dwarf galaxies, one must assume a functional form for the DM density profile. Deriving the appropriate ULDM profile requires understanding the formation of DM halos and their non-linear evolution, a process too complex to be captured analytically and therefore studied through numerical simulations. For our purposes, we adopt and combine results from simulations in the literature.

In this section, we describe the DM profiles used in our analysis. We first introduce the Navarro–Frenk–White (NFW) profile~\cite{Navarro:1995iw, Zhao:1995cp,Hernquist:1990be}, which describes the distribution of CDM in halos. We then present the soliton solution~\cite{Schive:2014dra} for ULDM halos. Finally, we describe the model that combines a central soliton core with an outer NFW-like profile.
 
A halo can be defined in various ways. Following \cite{Goldstein:2022pxu}, we define it as the spherical region of radius $r_{200}$ within which the mean enclosed density is $\Delta=200$ times the mean density of the universe, and we denote the mass enclosed within this region by $M_{200}$.

\subsection{Outer Halo Envelope: the NFW profile}
The NFW profile~\cite{Navarro:1995iw}, along with its generalized form~\cite{Zhao:1995cp,Hernquist:1990be}, was originally identified in N-body simulations of CDM, in which dark matter particles are assumed to have negligible initial thermal velocities and therefore no appreciable suppression of gravitational collapse at small scales. These simulations further showed that the resulting CDM halos are well described by a nearly universal density profile. In this work, we use the NFW profile to describe the outer region of the halo, adopting the original form of \cite{Navarro:1995iw},

\begin{equation}
\rho_{\mathrm{NFW}}(r)=\frac{\rho_s}{\left(r / r_s\right)\left[1+\left(r / r_s\right)\right]^{2}}~,
\label{eq: NFW}
\end{equation}
where $\rho_s$ is the characteristic density and $r_s$ is the scale radius. In the outer region, $r \gg r_s$, the density profile falls off as $\sim r^{-3}$, matching the scaling of the outer region in the ULDM simulations.

\subsection{Soliton Core Profile}

Because the collapse of ULDM halos is a highly non-linear process, numerical simulations are required to determine their density profiles. A number of such simulations have been performed~\cite{Chan:2021bja,Lopez-Sanchez:2025osk,Mocz:2017wlg,Chen:2026dst,Nori:2020jzx,Yavetz:2021pbc,May:2021wwp,May:2022gus,Mina:2020eik,Woo_2009,Schive:2014dra,Liao:2024zkj,Schwabe:2021jne,Elgamal:2023yzt,Teodori:2025rul,Schwabe:2016rze}, including cosmological simulations of large-scale structure formation (see, e.g., \cite{Woo_2009,Schive:2014dra,May:2021wwp,May:2022gus,Liao:2024zkj,Chen:2026dst,Elgamal:2023yzt}) and cosmological zoom-in simulations that resolve the internal structure of individual halos (see, e.g., \cite{Nori:2020jzx,Mina:2020eik,Schwabe:2021jne,Elgamal:2023yzt}). Complementary studies employ idealized halo-merger simulations (see, e.g., \cite{Mocz:2017wlg,Amin:2022pzv,Yavetz:2021pbc,Teodori:2025rul,Lopez-Sanchez:2025osk,Schwabe:2016rze}). These simulations show that a ULDM halo can develop a solitonic core at its center, which transitions smoothly into an NFW-like envelope at larger radii, consistent with the density profiles found in collisionless CDM simulations. These solitonic cores are a consequence of the finite Jeans length of ULDM on nonlinear scales, and correspond to the ground-state solution of the Schrödinger-Poisson equations. Below, we first introduce the scalar ULDM soliton profile, parameterized based on some of these simulations, and then generalize it to  higher-spin ULDM.

\subsubsection{Scalar DM Soliton Profile}
We adopt the soliton core parameterization from the work~\cite{Schive:2014hza} which performed one of the first cosmological simulations of cold, wavelike DM. For spin-0 scalar ULDM halos at the present epoch, the core density profile can be analytically written as:
\begin{equation}
\rho_{\text {soliton }}(r)=\frac{1.9\left(10 m_{22}\right)^{-2}\left(r_c / \mathrm{kpc}\right)^{-4}}{\left[1+9.1 \times 10^{-2}\left(r / r_c\right)^2\right]^8} 10^9 M_{\odot} \mathrm{kpc}^{-3}~,
\label{eqn: scalar profile}
\end{equation}
where the scalar particle mass $m$ is scaled as $m_{22} \equiv m/(10^{-22}$ eV$)$, $M_\odot$ is the solar mass, and $r_c$ is the soliton core radius, defined as the radius at which the soliton density falls to half of its central value, $\rho_{\text{soliton}}(r \rightarrow 0)$. This expression is accurate to within $2\%$ throughout the interval $0<r \lesssim 3 r_c$~\cite{Schive:2014hza}. 

The simulations in~\cite{Schive:2014hza} also predicted the scaling relation $M_{\rm core} \propto (E_h/M_h)^{1/2}$. Here, $M_{\rm core}$ is the mass of soliton core in the halo,\footnote{The definition of core mass varies across studies, for instance,~\cite{Amin:2022pzv} defines it as the soliton mass enclosed within $2r_c$, whereas~\cite{Schive:2014hza,Du:2016zcv} uses the mass within $r_c$. This difference does not affect the halo-mass exponent, $\gamma$, in the core-halo mass relation.} $E_h$ is the total halo energy and $M_h$ is the mass of the halo (in our study, we take $M_{h}$ to be $M_{200}$. We will use them interchangeably below). Under virial condition in the spherical collapse model, the scaling relation yields core-halo mass relation, $M_{\rm core}\propto M_{200}^\gamma$ with $\gamma=1/3$. Thus, the core radius which is inversely proportional to $M_{\rm core}$ is given by:
\begin{equation}
r_c \approx 1.5 m_{22}^{-1}\left(\frac{M_{200}}{10^9 M_{\odot}}\right)^{-1 / 3} \mathrm{kpc}~.
\label{eqn: rc}
\end{equation}
The ULDM soliton core profile for a scalar particle is therefore parameterized entirely by the halo mass and the DM particle mass parameter, $\left\{M_{200},m_{22}\right\}$. 

Although we take the~\cref{eqn: rc} as the benchmark parameterization for scalar ULDM profile, subsequent works 
have argued that the relation $M_{\rm core} \propto (E_h/M_h)^{1/2}$ may not be fundamental to the system~\cite{Schwabe:2016rze}. Under the Schrödinger–Poisson scaling symmetry, both sides scale linearly with the scaling parameter. Their proportionality therefore follows from the symmetry itself and does not point to a fundamental physical origin. A more fundamental formulation is given by relations between scale-invariant quantities, namely the core-to-halo mass ratio $M_{\rm core}/M_h$ and the dimensionless parameter, $\Xi = (E_h/M_h^3)/(Gm/\hbar)^2$. Some simulation works of soliton and halo mergers~\cite{Amin:2022pzv,Mocz:2017wlg} find $M_{\rm core}/M_h \propto \Xi^{1/3}$, which
corresponds to $M_{\rm core} \propto M_h^{5/9}$. However, these core-halo relations are still in debate within cosmological simulations~\cite{Chan:2021bja,May:2021wwp} and thus, there is no consensus on the
core--halo relation even for scalar ULDM at the moment (see Sec.~4.2.1 of Ref.~\cite{Chan:2021bja} for a summary of existing work). Furthermore,~\cite{Chan:2021bja}  argued that the dispersion in core-halo mass relation is potentially due to different formation and merger histories of the halos.

\subsubsection{Generalizing to Higher-spin DM Core Profile}
\label{subsec: vdm profile}
\begin{figure}[ht]
    \begin{center}\includegraphics[width=0.6
    \textwidth]{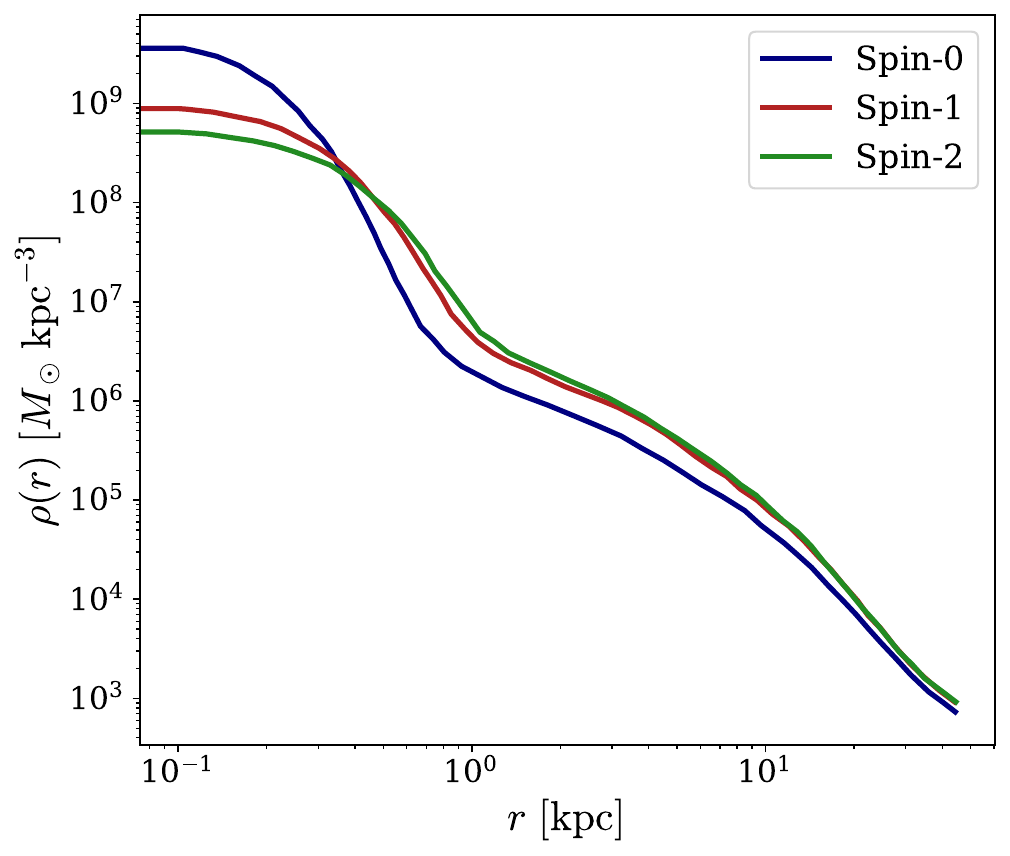}
    \end{center}
    \caption{Soliton density profiles for ULDM particles of different spins, obtained by spherically averaging the simulations performed in \cite{Lopez-Sanchez:2025osk}. These simulations were carried out with varying numbers of solitons, and each profile is the resulting radial average. 
  \label{fig: soliton profile} }
		\end{figure}

Beyond spin-0 ULDM, Refs.~\cite{Amin:2022pzv,Lopez-Sanchez:2025osk} have also performed numerical simulations of bosonic ULDM with non-zero spin, e.g., spin-1 vector DM and even spin-2 bosonic DM, using identical initial conditions. As in the scalar case, a soliton-like core forms at the halo center and again transitions into an $r^{-3}$-like tail in the outer region of the halo. Several important features nonetheless distinguish the scalar and higher-spin cases. In this section, we focus on the differences in the core; the transition between the core and outer regions is discussed in the next section. For a given set of initial conditions, e.g., a fixed number or fixed masses of initial solitons, the core radius $r_c$ is larger for higher-spin ULDM than for the scalar case, and the ratio of the core mass to the total halo mass is lower than that for scalar DM \cite{Amin:2022pzv,Lopez-Sanchez:2025osk}. In other words, starting from the same initial conditions, bosonic ULDM with non-zero spin tends to form a more diffuse core than in the scalar case, as illustrated in the schematic picture in \cref{fig: soliton profile}.

With these differences in mind, we propose a generalized form of the higher-spin core density profile, parametrized  by two additional parameters:
\begin{itemize}
    \item An overall normalization factor, \norm, to account for the change in the central density of a higher-spin DM halo relative to the scalar DM case.
    \item A core-halo mass relation factor, \rcnorm, to account for deviations of the halo-mass exponent from its value in the scalar ULDM case.
\end{itemize}
The generalized ULDM core density profile, which can accommodate multiple bosonic 
ULDM scenarios with different spins, is then given by:
\begin{equation}
\Tilde{\rho}_{\text {soliton }}(r)=\mu \times \frac{1.9\left(10 m_{22}\right)^{-2}\left(\Tilde{r}_c / \mathrm{kpc}\right)^{-4}}{\left[1+9.1 \times 10^{-2}\left(r / \Tilde{r}_c\right)^2\right]^8} 10^9 M_{\odot} \mathrm{kpc}^{-3}~,
\label{eqn: VDM soliton}
\end{equation}
where scale radius, $\Tilde{r}_c$ is now given by:

\begin{equation}
    \Tilde{r}_c \approx 1.5 m_{22}^{-1}\left(\frac{M_{200}}{10^9 M_{\odot}}\right)^{-\left( \frac{1}{3}-\psi \right)} \mathrm{kpc}~.
    \label{eqn: vector rc}
\end{equation}
In this parametrization, $M_{\rm core} \sim \mu m_{22}^{-2} \tilde{r}_c^{-1} \sim \mu m_{22}^{-1}M_{200}^{(1/3-\psi)}$. 
Thus, now the exponent in the core-halo mass relation is given by $\gamma=1/3-\psi$. The scalar DM case based on Ref.~\cite{Schive:2014hza} is recovered for \norm~$=1$ and \rcnorm~$=0$, for which the core density profile reduces to \cref{eqn: scalar profile}. 
For a set of identical initial conditions, higher-spin DM with $s > 0$ leads to a lower central density; in our
parametrization this is encoded by $\mu < 1$. The exponent $\psi$ instead describes how the core scales with the halo mass. 
Simulations that report the core--halo relation for vector ULDM find an exponent steeper
than the reference value 1/3: $\gamma\simeq0.45$ corresponding to $\psi \simeq -0.12$ from soliton merger simulation at present epoch $z=0$~\cite{Amin:2022pzv}, or $\gamma \simeq 0.64$ at high redshift $z \approx 790$,
corresponding to $\psi \simeq -0.3$~\cite{Chen:2026dst}.  
Thus, taking $\gamma=1/3$ as the benchmark, the expected sign for higher-spin ULDM is $\psi < 0$.
We note that because \cref{eqn: vector rc} is written relative to a reference mass of
$10^9\,M_\odot$, the ratio $\tilde r_c / r_c = (M_{200}/10^9\,M_\odot)^{\psi}$ exceeds
unity below $10^9\,M_\odot$ and falls below one above $10^9\,M_\odot$ for any $\psi < 0$. A non-zero
$\psi$ thus changes the core radius in opposite directions on either side of
$10^9\,M_\odot$. The diffuseness of the core at a given halo mass is controlled by
$\mu$, not by $\psi$ alone.

If the scalar relation is instead
$\gamma = 5/9$ as discussed in the previous section, the scalar point shifts to $\psi = -2/9$, and the constraints on $\psi$
reported in \cref{sec:result} should be read relative to that value. The relative difference in the exponent $\Delta \psi$ between scalar and vector ULDM would shift as well. We also want to highlight that
$\Xi$-based relation discussed in the previous section also modifies the dependence of the core mass on the particle mass
(see eq.~(15) of Ref.~\cite{Chan:2021bja}). We do not model this additional freedom
here, and note that at fixed $m_{22}$, modified exponent of $m_{22}$ is partially degenerate with the normalization
$\mu$.

Finally, we want to comment on the differences between our parametrization and earlier work on generalizing the scalar DM soliton profile beyond eq.~\eqref{eqn: scalar profile}. Ref.~\cite{Chan:2021bja} proposes a parameterization to capture the intrinsic diversity in core-halo relation found in their simulation of scalar ULDM. An analysis using this parameterization was subsequently performed in Ref. \cite{Wardana:2026qag}, using stellar line-of-sight velocities of Milky Way dwarf spheroidal galaxies. We emphasize that our phenomenological parameterization does not consider varying the scaling of $m_{22}$ shown in their work. Instead, it is motivated by the aim of capturing the features of higher-spin ULDM profiles differing from the scalar ULDM case, with the minimum number of additional parameters. The two parametrization with different motivations are not degenerate, though both adopt more free model parameters some of which are similar, and lead to quantitative different results.

\subsection{Combined Halo Profile}
\label{subsec: halo profile}
Following \cite{Robles:2018fur}, we model the complete halo density profile by joining the inner soliton core of \cref{eqn: VDM soliton} to the NFW profile at a transition radius, $r_\alpha=\alpha r_c$:
\begin{equation}
\rho_{\mathrm{halo}}(r)= \begin{cases} \Tilde{\rho}_{\mathrm{soliton}}(r) & 0 \leq r \leq r_\alpha \\ \rho_{\mathrm{NFW}}(r) & r_\alpha \leq r \leq r_{200}\end{cases}~.
\label{eq:combined}
\end{equation}
At the transition radius, we impose continuity of the density, 
\begin{equation}
\tilde{\rho}_{\mathrm{soliton}}\left(r_\alpha\right)=\rho_{\mathrm{NFW}}\left(r_\alpha\right)~,
\end{equation}
and we require mass conservation within $r_{200}$,
\begin{equation}
\begin{aligned}
M_{200} & =4 \pi \int_0^{r_{200}} \rho_{\mathrm{halo}}\left(r^{\prime}\right) r^{\prime 2} d r^{\prime} \\
& =4 \pi \int_0^{r_\alpha} \tilde{\rho}_{\mathrm{soliton}}\left(r^{\prime}\right) r^{\prime 2} d r^{\prime}+ 4 \pi \int_{r_\alpha}^{r_{200}} \rho_{\mathrm{NFW}}\left(r^{\prime}\right) r^{\prime 2} d r^{\prime}~.
\end{aligned}
\end{equation}
Note that the combined profile introduces a further parameter, $\alpha$, characterizing the transition between the core and NFW regions. For scalar ULDM, simulations showed that it is about $\alpha \approx 3$ in dwarf-sized galaxies~\cite{Robles:2018fur}. Its value may vary with the spin of the ULDM~\cite{Lopez-Sanchez:2025osk}.  


In our analysis, we perform Bayesian inference using the {\sc MultiNest}~\cite{Feroz:2008xx} algorithm through its Python interface, {\sc PyMultiNest}~\cite{2016ascl.soft06005B}. {\sc MultiNest} implements nested sampling, a Monte Carlo technique that efficiently explores the parameter space by iteratively sampling within regions of increasing likelihood, while also providing an estimate of the Bayesian evidence. The parameters of our analysis and their uniform priors are listed in~\cref{tab: parameters}. Note that the first three parameters already appear in the analysis of scalar ULDM~\cite{Goldstein:2022pxu}, while the remaining three are new here.

\begin{table}
\centering
\resizebox{\textwidth}{!}{%
\begin{tabular} {| c| c| c|}
\hline
 \textbf{Parameter} & \textbf{Description} & \textbf{Prior}\\
\hline
$\beta_a$ & Orbital anisotropy~ (\cref{eq:beta}) & $-1 \leq-\log _{10}\left(1-\beta_a\right) \leq+1$ \\
\hline
$m_{22}$ &Scaled ULDM particle mass & $-1 \leq \log _{10}\left(m_{22}\right) \leq 4~$ \\
\hline
$M_{200}$ & Halo mass & $ \log _{10}\left(M_{200}^{\min }\left(m_{22}\right) / M_{\odot}\right) \leq \log _{10}\left(M_{200} / M_{\odot}\right)$ \\
& & $\leq \log _{10}\left(1 \times 10^{11}\right)$ \\
\hline
$\mu$ & Soliton core density scaling~(\cref{eqn: VDM soliton}) & $-2 \leq \log _{10}(\mu) \leq 2$ \\
\hline
$\psi$ & Core-halo mass exponent factor~(\cref{eqn: vector rc}) & $-3 \leq \log _{10}(|\psi|) \leq \log _{10}(2)$ \\
\hline
$\lambda\equiv3-\alpha$ & Transition radius scaling~(\cref{eq:combined}) & $-3 \leq \log _{10}(|\lambda|) \leq \log _{10}(2)$ \\
\hline
\end{tabular}%
}
\caption{Parameters in our analysis and their priors. $M_{200}^{\min}(m_{22})$ is the minimum allowed halo mass for a given $m_{22}$. \label{tab: parameters}}
\end{table}

Because of quantum pressure, a ULDM core of a given size cannot be hosted by an arbitrarily small halo. Therefore, for each value of $m_{22}$, there exists a minimum halo mass $M_{200}^{\rm min}$ below which no stable halo can exist. It is determined by solving the following equation~\cite{Goldstein:2022pxu}:
\begin{equation}
\begin{aligned}
& \frac{M_{200}^{\min }-\left.M_{\text {soliton }}\right|_{r_\alpha}}{\left.4 \pi r_\alpha^3 \tilde{\rho}_{\text {soliton }}\right|_{r_\alpha}}=\left(1+\frac{r_\alpha}{r_s}\right)^2\left(\frac{r_s}{r_\alpha}\right)^2 
\times\left[\frac{r_s\left(r_\alpha-r_{200}\right)}{\left(r_{200}+r_s\right)\left(r_\alpha+r_s\right)}+\log \left(\frac{r_{200}+r_s}{r_\alpha+r_s}\right)\right]~,
\end{aligned}
\end{equation}
where $\tilde{\rho}_{\text {soliton }}$ is the generalized ULDM soliton density profile given by \cref{eqn: VDM soliton} and $M_{\text {soliton }}$ is the corresponding soliton core mass:
\begin{equation}
M_{\text {soliton }}=\int_0^{r_\alpha} \tilde{\rho}_{\text {soliton }}(r^\prime) d^3 r'.
\end{equation}

Among all the priors, the prior of $\beta_a$ follow~\cite{Goldstein:2022pxu}. The upper limit on the halo mass, $M_{200}=10^{11}M_\odot$, is set by the allowed mass range for dwarf galaxies, and we consider a broad range of $m_{22}$ spanning five orders of magnitude. For the two additional parameters of the generalized ULDM profile, $\psi$ and $\mu$, we adopt symmetric logarithmic priors centered at $\psi=0$ and $\mu=1$, respectively, corresponding to their values in the scalar ULDM limit in~\cite{Schive:2014hza}. For dwarf-sized halos ($M_{200}\sim [10^8 - 10^{10}]M_{\odot}$), the transition radius parameter in the scalar ULDM case is approximately $\alpha \approx 3$~\cite{Robles:2018fur}. We introduce an equivalent parameter, $\lambda$, to characterize deviations from this reference value, such that $\alpha=3-\lambda$. We then impose a symmetric logarithmic prior around $\lambda=0$.
To avoid numerical convergence issues, we impose a lower cutoff of $10^{-3}$ for both $|\psi|$ and $|\lambda|$, sufficiently close to zero to approximate the scalar limit. We also choose broad ranges of $\mu, |\psi|, |\lambda|$, covering profiles deviating significantly from the scalar one.     
Imposing these priors on $\psi$, $\mu$ and $\lambda$ allow us to investigate whether the stellar kinematic data favor the more diffused core profiles associated with non-zero-spin ULDM over those predicted in the scalar case.

\section{Results and Discussion}
\label{sec:result}

\subsection{Observational Data}

Our analysis follows the same methodology and datasets as \cite{Goldstein:2022pxu}. We perform our analysis using stellar kinematic data from six classical dwarf spheroidal galaxies: Fornax, Sculptor, Draco, Sextans, Ursa Minor, and Carina. For Fornax, Carina, Sculptor, and Sextans, we adopt the datasets from \cite{2009AJ....137.3100W}. A detailed description of the target selection, observational methods, and data reduction procedures can be found in \cite{2007ApJ...667L..53W}. 

For Draco and Ursa Minor, we use the catalogs compiled by \cite{Spencer_2018}, which combine data from multiple literature sources spanning 30 years of observations. For stars with multiple observations, measurements are combined into a single velocity value by taking the mean, weighted by the inverse square of the corresponding velocity errors.

We use the membership probabilities provided in \cite{Walker:2008fc} to identify member stars of each dwarf galaxy and to distinguish them from foreground contamination due to Milky Way. These probabilities are derived under the assumption that the velocity dispersion remains constant with projected galactocentric distance. We only select stars with membership probability $> 95 \%$. The resulting samples contain 774, 2483, 1365, and 441 probable members in Carina, Fornax, Sculptor, and Sextans, respectively. For Draco, there are 692 probable members (341 of which have multiple observations), and for Ursa Minor, there are 680 probable members (284 with multiple observations). The half-light radii for all galaxies are taken from \cite{1995MNRAS.277.1354I}.

\subsection{Results}

\begin{table}
\centering
\small
\setlength{\tabcolsep}{4pt}
\begin{tabular}{|l|c|c|c|c|c|c|}
\hline
 & \textbf{Fornax} & \textbf{Carina} & \textbf{Sculptor} & \textbf{Sextans} & \textbf{Draco} & \textbf{Ursa Minor} \\
\hline
{\boldmath$\log_{10}(1-\beta_a)$} & $> 0.316$ & $0.53^{+0.46}_{-0.40}$ & $0.55^{+0.42}_{-0.33}$ & $0.40^{+0.48}_{-0.35}$ & $0.37^{+0.56}_{-0.33}$ & $> 0.230$ \\
\hline
{\boldmath$\log_{10}(M_{200}/M_\odot)$} & $8.53^{+0.43}_{-0.29}$ & $8.65^{+0.74}_{-0.66}$ & $9.06^{+0.62}_{-0.56}$ & $8.59^{+1.4}_{-0.95}$ & $9.66^{+1.1}_{-0.87}$ & $8.62^{+0.55}_{-0.45}$ \\
\hline
{\boldmath$\log_{10}(m_{22})$} & $1.98^{+1.3}_{-0.48}$ & $> 2.12$ & $2.46^{+1.2}_{-0.62}$ & $2.68^{+1.1}_{-0.93}$ & $2.7^{+1.2}_{-1.0}$ & $> 1.83$ \\
\hline
{\boldmath$\log_{10}(\mu)$} & $0.94^{+0.54}_{-1.9}$ & $0.08^{+0.85}_{-1.4}$ & $0.04^{+0.79}_{-1.3}$ & $0.0^{+1.3}_{-1.7}$ & $-0.7^{+1.1}_{-1.2}$ & $0.01^{+0.94}_{-0.96}$ \\
\hline
{\boldmath$\psi$} & $-0.01^{+0.17}_{-0.22}$ & $-0.02^{+0.24}_{-0.38}$ & $-0.01^{+0.42}_{-0.45}$ & $-0.01^{+0.23}_{-0.27}$ & $0.00^{+0.23}_{-0.27}$ & $0.01^{+0.51}_{-0.41}$ \\
\hline
{\boldmath$\lambda$} & $0.01^{+0.36}_{-0.29}$ & $0.00^{+0.43}_{-0.39}$ & $0.00^{+0.41}_{-0.35}$ & $-0.03^{+0.45}_{-0.50}$ & $0.00^{+0.33}_{-0.34}$ & $0.04^{+0.80}_{-0.56}$ \\
\hline
\end{tabular}
\caption{95\% C.L. constraints and ranges on the DM halo profile model proposed in \cref{subsec: halo profile}, using stellar velocity dispersion data for six dwarf galaxies: Fornax, Carina, Sculptor, Sextans, Draco, and Ursa Minor. }
\label{tab:constraints}
\end{table}

\begin{figure}[H]
\centering
\includegraphics[scale=0.5]{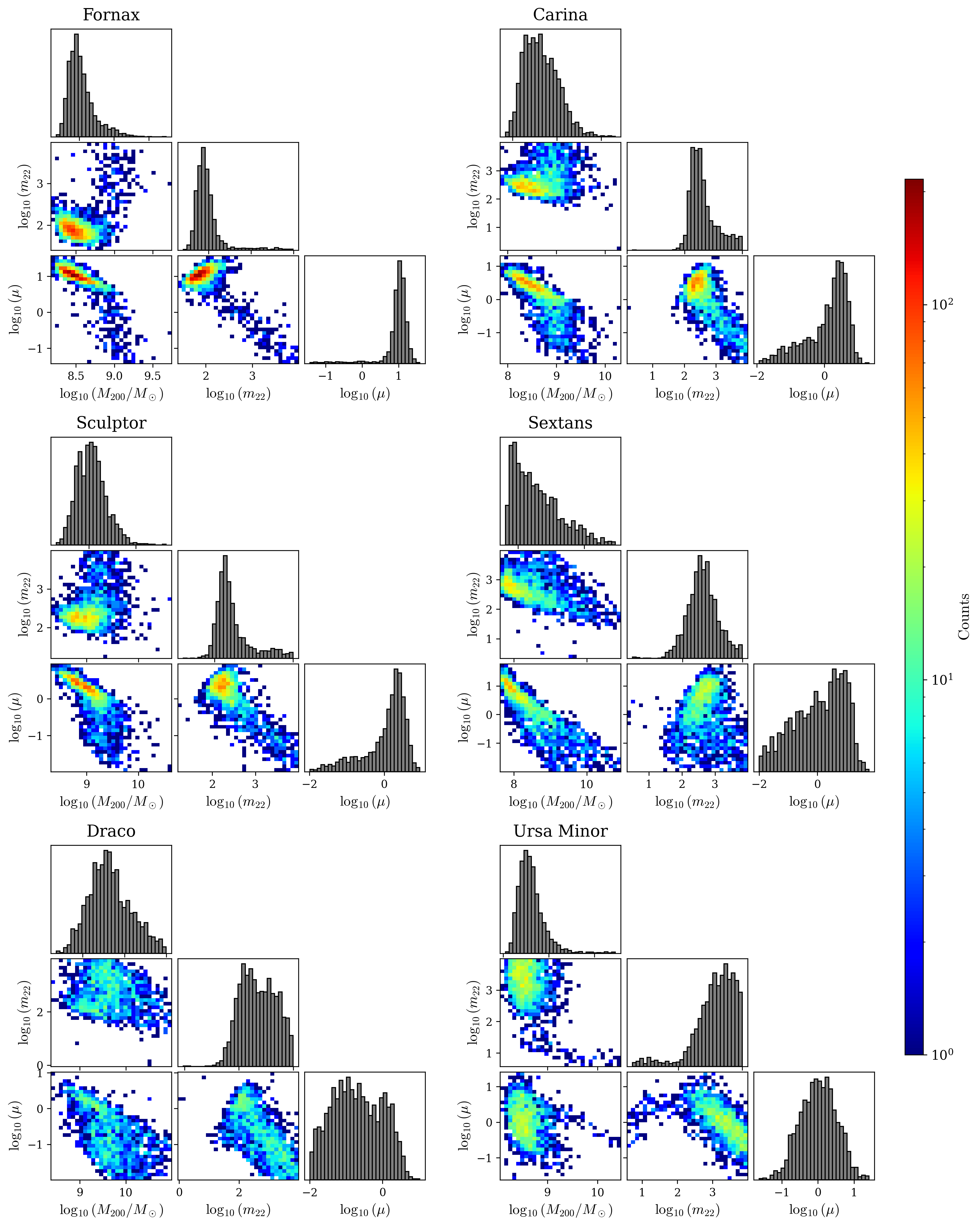}
\caption{Posterior distributions of the scaled ULDM particle mass, $m_{22}$, the core central density parameter $\mu$, and halo mass $M_{200}$, for the  generalized ULDM halo profile model described in \cref{subsec: halo profile}, using stellar kinematic data from all six dwarf galaxies.
\label{fig:m-M-mu}}
\end{figure}

\begin{figure}[H]
\centering
\includegraphics[scale=0.5]{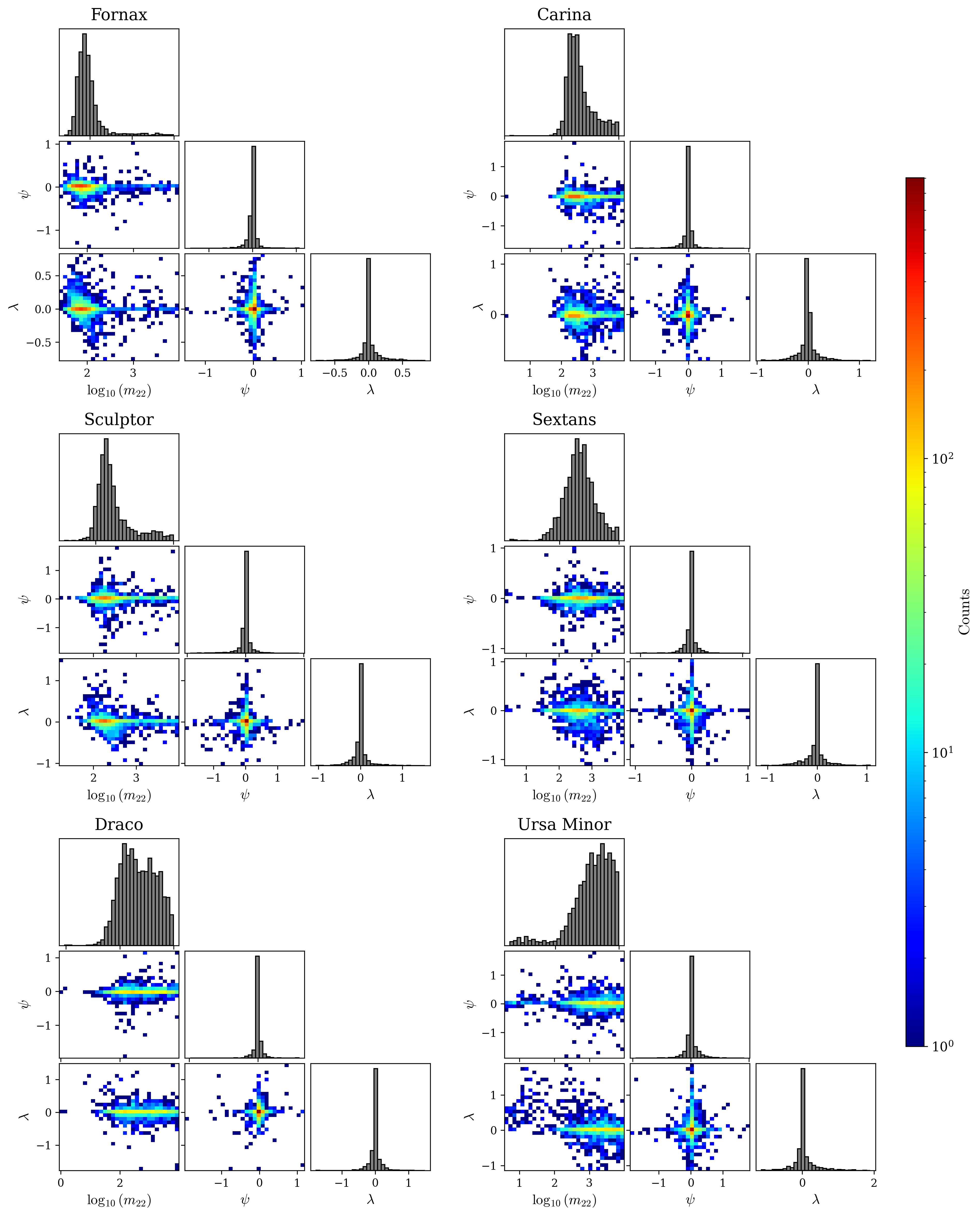}
\caption{Posterior distributions of the scaled ULDM particle mass, $m_{22}$, core-halo relation parameter, $\psi$ and the transition radius parameter, $\lambda$, for the generalized ULDM halo profile model described in \cref{subsec: halo profile}, using stellar kinematic data from all six galaxies.
\label{fig:m-psi-lambda}}
\end{figure}

\begin{figure}[H]
\centering
\includegraphics[scale=0.5]{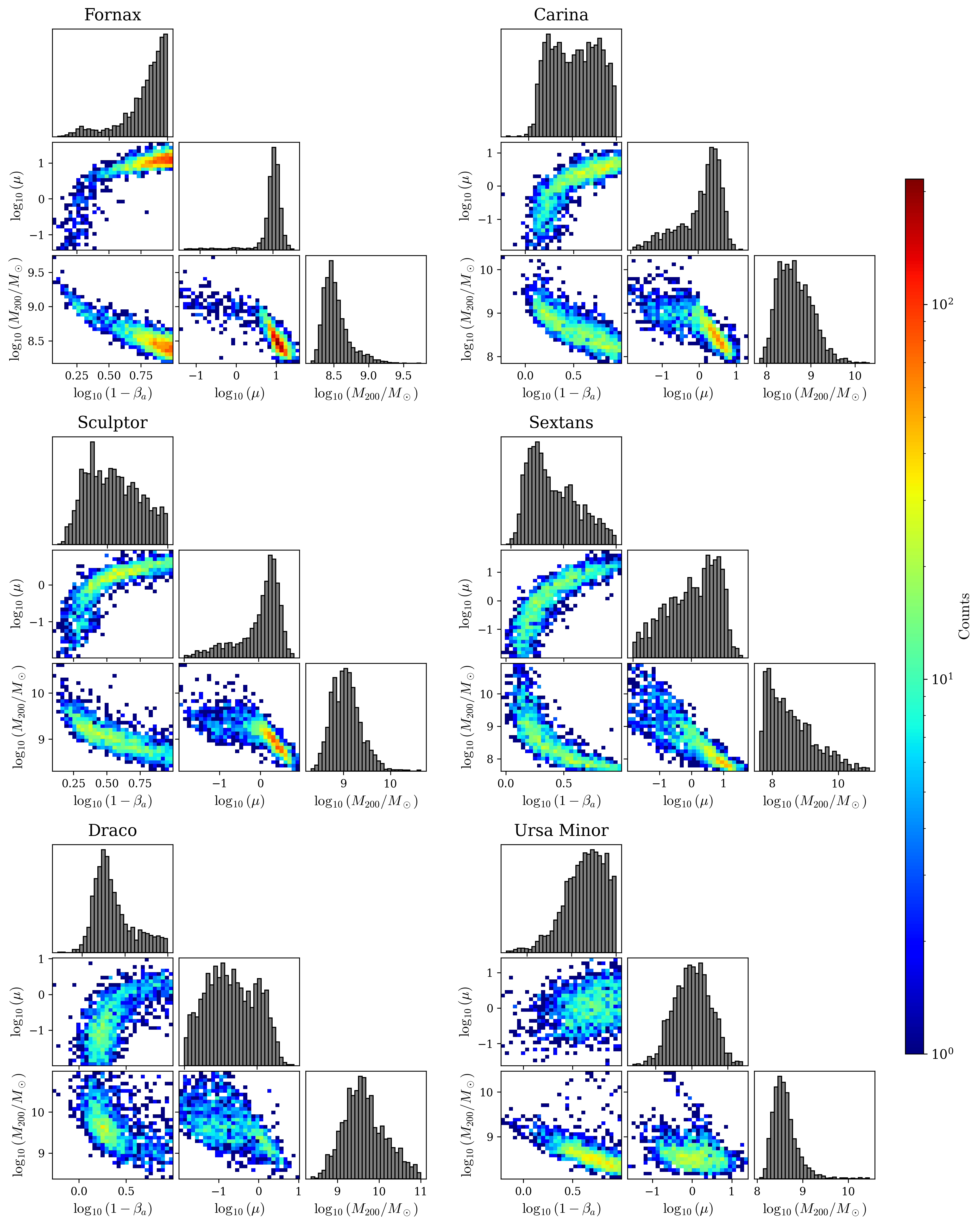}
\caption{Posterior distributions of the scaled ULDM central density parameter, $\mu$, halo mass $M_{200}$, and the orbital anisotropy parameter $\beta_a$, for the  generalized ULDM halo profile model described in \cref{subsec: halo profile}, using stellar kinematic data from all six dwarf galaxies.
\label{fig:m-M-beta}}
\end{figure}

Our main results are presented in \cref{tab:constraints}, which lists the 95\% C.L. ranges and constraints on the six generalized DM halo profile parameters for the six galaxies. The corresponding posterior distributions are shown in \cref{fig:m-M-mu} for the subset $\{ M_{200}, m_{22}, \mu \}$, in \cref{fig:m-psi-lambda} for  $\{m_{22}, \psi, \lambda\}$, for $\{\beta_a, M_{200}, \mu \}$ in~\cref{fig:m-M-beta}. Fig.~\ref{fig:result1}--\cref{fig:result6} in \cref{app: full posteriors} show the posterior distributions of all the free parameters in our analysis. For the scalar ULDM (a special case of the generalized ULDM profile with $\psi=\lambda=0$ and $\mu=1$), we refer the reader to \cite{Goldstein:2022pxu}, where the posterior distributions for the model parameters $\{M_{200}, m_{22}, \beta_a\}$ are presented in figs. 2 and 3. 


From our results, for all galaxies, the host halo mass 
always peaks in region $\sim [10^{8.5} - 10^{9.5}]\,M_{\odot}$, while the particle mass is 
constrained to $m_{22} \gtrsim 30-130$ at the $95\%$ 
C.L., depending on the galaxy 
(\cref{tab:constraints}). Similar results for $M_{200}$ and $m_{22}$ appear in previous work on scalar ULDM with $\mu, \psi$ and $\alpha$ held fixed~\cite{Goldstein:2022pxu}, 
where the halo mass is $\sim [10^8 - 10^9] M_\odot$ and 
$m_{22} \gtrsim 100$.  The additional parameters $\mu$, 
$\psi$, and $\alpha$ introduced here therefore do not 
change this conclusion qualitatively. 

As shown in \cite{Goldstein:2022pxu}, $M_{200}$ and $m_{22}$ are anticorrelated for the scalar DM profile, with two distinct regions in the two-dimensional parameter space (one at lower halo masses around $M_{200} \sim 10^9 M_\odot$, and one at  higher masses, $M_{200} \gtrsim 10^{10} M_\odot)$.  For higher-spin ULDM, however, \cref{fig:m-M-mu} shows that while the $M_{200}-m_{22}$ anticorrelation persists, it is weaker than in the scalar ULDM case, and the two-region posterior is replaced by a single preferred region in parameter space. For Draco and Ursa Minor, the anticorrelation is no longer apparent and effectively disappears. In the scalar DM case, this degeneracy arises because the Jeans equation is most sensitive to the enclosed mass near the Plummer radius \cite{Strigari:2008ib}, so any combination of $M_{200}$ and $m_{22}$ yielding a similar enclosed mass at this radius fits the kinematic data equally well. As the particle mass increases, quantum pressure effects weaken and the profile approaches the cuspy, classical NFW-like limit. Since a cusped profile concentrates more mass toward the center than a cored one at fixed $M_{200}$, fitting the inner data points requires a larger $M_{200}$ for more extended (lower-$m_{22}$) cores, while more cusped (higher-$m_{22}$) profiles can match the data with a smaller $M_{200}$.

The weakening of the $M_{200}-m_{22}$ anticorrelation in the generalized ULDM case is likely due to the additional profile parameters, mainly the central core density, $\mu$, which introduce further degeneracies into the fits. As shown in \cref{fig:m-M-mu}, $M_{200}$ is anticorrelated with $\mu$: profiles with higher central densities (larger $\mu$) yield the enclosed mass needed to fit the inner data points at a smaller $M_{200}$, while those with lower central densities require a larger $M_{200}$ to compensate.

The constraints on $\mu$ are broad, spanning roughly two orders of magnitude at 95\% C.L. in every
galaxy (\cref{tab:constraints}), and the scalar value $\mu = 1$ lies within the 95\%
range for all six galaxies. The posterior modes nevertheless differ from galaxy to galaxy: the
three best-measured systems, Fornax, Carina and Sculptor, peak at $\log_{10}\mu \sim
0.5$--$1$, Sextans and Ursa Minor peak near $\mu \simeq 1$, and Draco peaks below it, at
$\log_{10}\mu \simeq -0.7$. In the $m_{22}$--$\mu$ plane of \cref{fig:m-M-mu}, the
posteriors for Fornax, Carina and Sculptor show an extended tail toward larger $m_{22}$
and smaller $\mu$, along which $M_{200} \sim 10^9\,M_\odot$ remains nearly constant.
This anticorrelation arises because larger values of $\mu$ compensate for the lower
central density at smaller $m_{22}$, whereas smaller values of $\mu$ suffice at larger
$m_{22}$ as the profile becomes increasingly cuspy.

The preference for $\mu > 1$ in Fornax, Carina and Sculptor has a direct physical
interpretation. For a soliton, the central density and core radius are not independent: the
scaling symmetry of the Schr\"odinger--Poisson system fixes $\rho_c \propto
m^{-2}r_c^{-4}$, which is the origin of the $(r_c/\mathrm{kpc})^{-4}$ prefactor in
\cref{eqn: scalar profile}. A profile with $\mu > 1$ at fixed $\tilde r_c$ is therefore
denser than any soliton of that radius, scalar or otherwise, and a preference for
$\mu > 1$ is a preference for a central mass concentration higher than ULDM cores
provide. In Fornax, where the mode lies at $\mu \simeq 9$, this is the clearest
indication in our analysis that the kinematic data favor a cusp-like profile. The direction
expected for higher-spin ULDM is the opposite, $\mu < 1$ (\cref{subsec: vdm profile}):
the central densities from simulation results of~\cite{Lopez-Sanchez:2025osk} in \cref{fig: soliton profile} correspond to $\log_{10}\mu
\simeq -0.6$ to $-0.8$.\footnote{Estimating the range of $\mu<1$ for ULDM with different non-zero spins may call for further studies.} Only Draco leans that way, and in no galaxy is such a
suppression required, although in no galaxy is it excluded either, given the width of
the posteriors. 


In \cref{fig:m-psi-lambda}, the posterior distribution of $\psi$ is  peaked
near $\psi \simeq 0$ for all six galaxies. This peak is largely a feature of the
adopted logarithmic prior, which is uniform in $\log_{10}|\psi|$ and therefore places
about 60\% of its weight at $|\psi| < 0.1$.
The information carried by the kinematic data is instead in the
width of the posterior: at 95\% C.L. we find $|\psi| \lesssim 0.2$--$0.5$, depending on
the galaxy (\cref{tab:constraints}), i.e. the data are consistent with the scalar
core--halo relation and show no preference for deviations from it. For reference, the
core--halo exponent $\gamma = 5/9$ found in some scalar simulations
\cite{Mocz:2017wlg,Amin:2022pzv} corresponds to $\psi = 1/3 - \gamma \simeq -0.22$,
which lies at the edge of the 95\% range for Fornax and within it for the other
galaxies. For vector ULDM, the exponent $\gamma \simeq 0.45$ obtained from soliton-merger simulations at $z=0$~\cite{Amin:2022pzv} corresponds to $\psi \simeq -0.12$, well within the 95\% C.L. range for Fornax. In contrast, the steeper relation $\gamma \simeq 0.64$ found at high redshift in Ref.~\cite{Chen:2026dst} corresponds to $\psi \simeq -0.31$, which lies outside the Fornax range. We note that the constraint on $\psi$ from any single galaxy is limited:
since \cref{eqn: vector rc} pivots at $M_{200} = 10^9\,M_\odot$, the effect of $\psi$ on
$\tilde r_c$ within one galaxy is degenerate with $m_{22}$ and is resolved only through
the width of that galaxy's $M_{200}$ posterior. A joint fit across halos spanning a
wider mass range is required to constrain the exponent directly, and we leave this to
future work. 

The posterior distribution of $\lambda$ in \cref{fig:m-psi-lambda} behaves in the same
way: it is  peaked at $\lambda \simeq 0$ ($\alpha \simeq 3$) in all six galaxies,
again largely as a consequence of the logarithmic prior. The 95\% ranges in
\cref{tab:constraints} correspond to $\alpha \approx 2$--$4$, consistent with the
values $\alpha \approx 2$--$5$ found in numerical simulations of both scalar and
higher-spin ULDM~\cite{Lopez-Sanchez:2025osk}. The weakness of this constraint reflects
the limited sensitivity of the stellar kinematics to the location of the soliton--NFW
transition, since the line-of-sight dispersion is governed mainly by the enclosed mass
within the half-light radius, which is dominated by the core.

Lastly, we comment on the orbital anisotropy parameter, $\beta_a$ in \cref{fig:m-M-beta}. In our parametrization, $\log_{10}(1-\beta_a)>0$ corresponds to tangentially biased stellar orbits.
The posteriors in~\cref{fig:m-M-beta} shows a clear positive correlation in $\log_{10}(1-\beta_a)-\mu$ plane, and an anticorrelation in $\log_{10}(1-\beta_a)-M_{200}$ plane respectively. Both follow from the mass--anisotropy degeneracy in the Jeans analysis. The observed line-of-sight velocity dispersion (see fig.~1 in~\cite{Goldstein:2022pxu}) is roughly flat from the inner region, $\mathcal{O}(0.01)$~kpc, out to about the half-light radius. If a larger fraction of the mass is concentrated within small radii below the half-light radius and the enclosed mass barely grows beyond the region of concentration, the line-of-sight velocity dispersion decreases with radius above the concentration region. A flat profile then requires either a less concentrated mass distribution or a tangential velocity bias, i.e.\ a larger $1-\beta_a$. 
Given roughly the same enclosed mass at half-light radius and same $M_{200}$, a cored profile with a larger $\mu$ makes the profile more centrally concentrated, and thus $1-\beta_a$ must increase to keep the profile of light-of-sight dispersion flat. This explains the $\log_{10}(1-\beta_a)-\mu$ positive correlation. Once $\mu$ and $M_{200}$ are both free, the requirement that the mass within the half-light radius remain fixed makes them anti-correlated as discussed above. This leads to anti-correlation between $\log_{10}(1-\beta_a)$ and $M_{200}$.

All six galaxies prefer tangentially biased orbits, with $\log_{10}(1-\beta_a) \simeq
0.4$--$0.55$ for Carina, Sculptor, Sextans and Draco, and one-sided constraints
$\log_{10}(1-\beta_a) > 0.32$ and $> 0.23$ for Fornax and Ursa Minor, respectively
(\cref{tab:constraints}). These values are consistent with those found for scalar ULDM
in fig.~3 of Ref.~\cite{Goldstein:2022pxu} over the same range of $M_{200}$. For Fornax,
however, the posterior rises monotonically to the upper edge of the prior,
$\log_{10}(1-\beta_a) = 1$, and this extreme anisotropy is reached together with the
large core normalization, $\mu \simeq 9$, and the low halo mass,
$M_{200} \simeq 10^{8.5}\,M_\odot$, preferred for that galaxy: the three parameters move
jointly along the mass--anisotropy degeneracy. We therefore regard the Fornax anisotropy
as limited by the prior rather than measured by the data, and note that the inferred
$M_{200}$ for Fornax is correspondingly sensitive to the freedom in $\mu$. A comparison
of the model line-of-sight dispersion profile with the binned data, which would
establish whether this behaviour reflects a genuine tension in the inner bins, is left
for future work.

\section{Conclusion}
\label{sec:con}
In this paper, we propose a generalized parametrization of the halo profile for ULDM with different spins. Specifically, we introduce three new parameters to capture deviations of the higher-spin ULDM profile from the widely-used scalar one. They account for variations in the normalization of the central core density, exponent in the core-halo mass relation, and transition radius from the core to the NFW outskirts. 

We then apply the stellar kinematics data from six classical dwarf galaxies to constrain this generalized parametrization. Bayesian analyses are conducted with logarithmic priors on the three new parameters in the generalized profile, as well as three conventional parameters: halo mass, orbital anisotropy and ULDM particle mass. 
The inferred halo masses lie in the range $10^{8.5}$--$10^{9.5}\,M_\odot$, and each
galaxy individually requires a particle mass $m \gtrsim 3\times10^{-21}$~eV; neither
result depends on the values of the new parameters. The core--halo exponent and the
transition radius are consistent with their scalar values, $|\psi| \lesssim 0.2$--$0.5$
and $\alpha \approx 2$--$4$ at 95\% C.L., with the  peaks of their posteriors at
the scalar point being largely a consequence of the logarithmic prior rather than a
preference of the data. The core-density normalization $\mu$ is consistent with the
scalar value in four galaxies, lies below it in Draco, and in Fornax prefers a core
denser than any soliton of the same radius, which we read as a preference for a cusped
profile. The factor-of-several suppression of the central density expected for
higher-spin ULDM is not required by any galaxy, but given the width of the posteriors it
is not excluded by any single galaxy either. The data therefore show no preference for
the more diffuse cores of higher-spin ULDM, without yet ruling them out.

Several limitations of the present analysis point to its natural extensions. First, each
galaxy is fitted independently. Since the core--halo exponent $\psi$ is a property of the
population rather than of a single halo, and since our parametrization pivots at
$M_{200} = 10^9\,M_\odot$ with most of the sample lying below it, the per-galaxy
constraints on $\psi$ arise only through the width of each galaxy's $M_{200}$ posterior.
A joint fit with shared particle parameters across halos spanning a wider mass range is
needed to constrain the exponent directly, and would also reduce the freedom of
individual galaxies, Fornax in particular, to trade core normalization against orbital
anisotropy. Second, the logarithmic priors adopted for $\psi$ and $\lambda$ concentrate
prior mass near the benchmark scalar point, so the  posterior peaks there should not be read
as evidence for the scalar profile; a reanalysis with flat priors, together with a
Bayesian evidence comparison between the generalized and scalar models, which is
straightforward since the latter is nested within the former, will be presented in
future work. Third, the scalar core--halo relation itself is not settled: simulations
report exponents between $\gamma = 1/3$ and $5/9$, which shifts the location of the
scalar reference point in $\psi$. In a similar vein, a direct comparison with higher-spin ULDM
requires a core--halo mass relation for vector and tensor fields at the present
epoch, which still needs further studies. Once such relations are confirmed, the spin-1 and spin-2 predictions can be placed directly in the
$(\mu, \psi)$ plane and tested against the kinematic data. Finally, our results are derived based on spherical Jeans modeling, subject to several well-known limitations. These include the assumption of spherical symmetry and dynamical equilibrium, as well as the degeneracy between the enclosed mass profile and the velocity anisotropy of the tracer population when only line-of-sight velocities are available. A number of mass-modeling techniques have been developed to address these issues. Among them are higher-order Jeans modeling~\cite{2026A&A...705A.212B}, Schwarzschild orbit-superposition modeling~\cite{2013MNRAS.433.3173B, 2019MNRAS.482.5241K}, distribution-function fitting~\cite{Wilkinson:2001ut, Kleyna:2001us, 2019MNRAS.482.1525V}, simulation-based inference~\cite{Nguyen:2022ldb, Nguyen:2025wjy}, and deep learning approaches~\cite{2023ApJ...942...26G, Exposito-Marquez:2022pqk}. Applying these methods to ULDM would allow a more robust assessment of the systematic uncertainties in our results. Perhaps the most promising avenue, however, is the inclusion of stellar proper motions. These provide direct access to the tangential velocity components and can therefore break the mass–anisotropy degeneracy. Refs~\cite{Vitral:2024rwl, Vitral:2025enn} have recently used proper motions to jointly constrain the dark matter density profiles and velocity anisotropies of Draco and Sculptor. Incorporating such three-dimensional kinematic data into ULDM analyses of dwarf spheroidals could potentially tighten the constraints, particularly in the inner regions where the predicted solitonic cores are most distinct.

\section*{Acknowledgements}
We thank Mustafa Amin for useful correspondence. P. S. and J. F. are supported by the NASA grant 80NSSC22K0811 and the DOE grant DE-SC-
0010010. S. M. K.  is supported by
National Science Foundation (NSF) Grant No. PHY-
2412666. This work was conducted using computational resources and services at the Center for Computation and Visualization, Brown University.

\appendix

\section{Full Posterior Distribution}
\label{app: full posteriors}
In this appendix, we show the full posterior distributions of all six parameters for the six dwarf galaxies we study. In all figures, $M_{200}$ is the halo mass, $m_{22}$ the scaled particle mass, $\beta_a$ the
orbital anisotropy, $\mu$ the central core density normalization, $\psi$ the offset of
the core--halo mass exponent, and $\lambda = 3 - \alpha$ the offset of the transition
radius parameter, as defined in \cref{subsec: vdm profile,subsec: halo profile}.

\begin{figure}[H]
\centering
\includegraphics[scale=0.55]{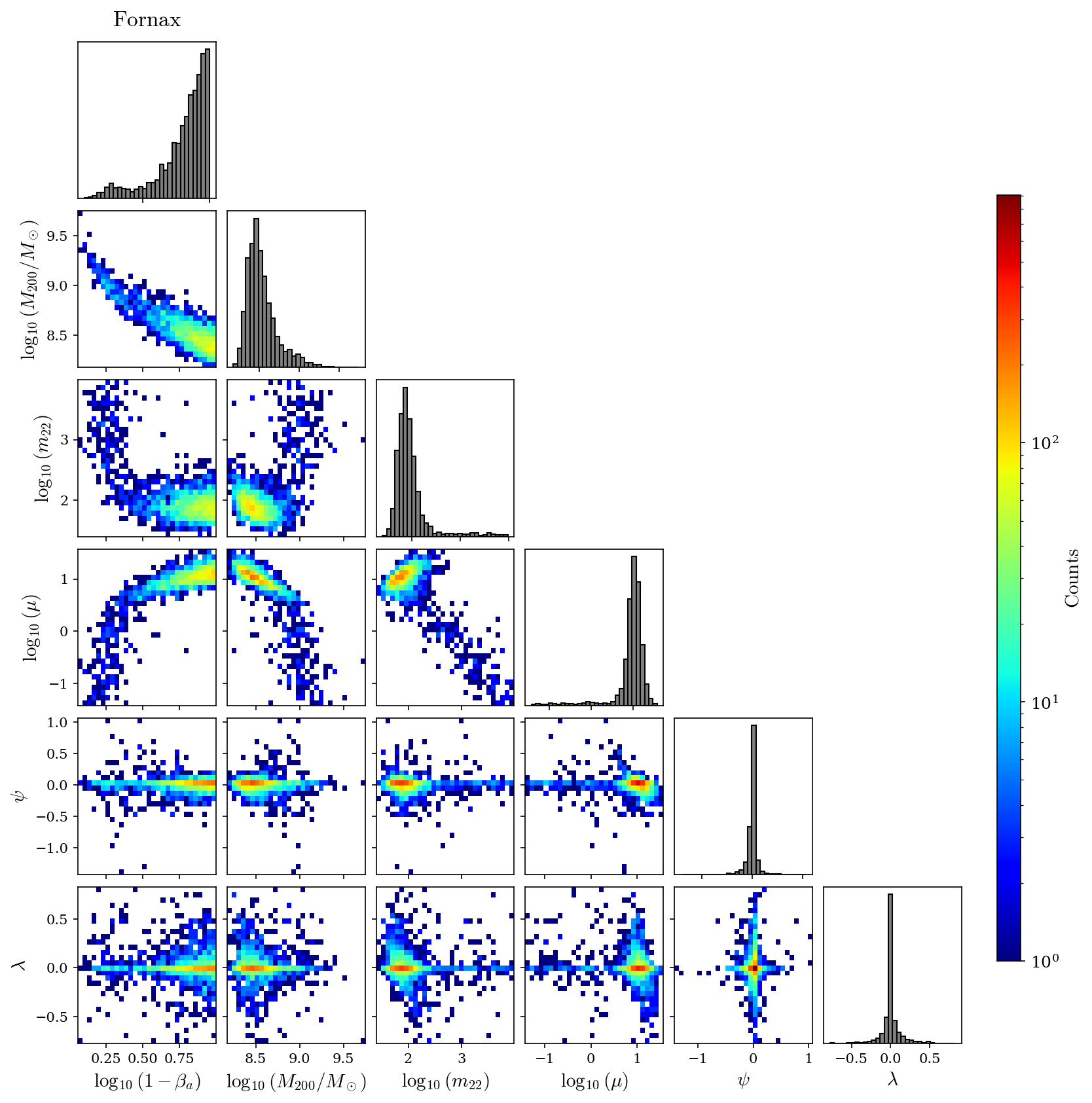}
\caption{Posterior distribution of spin-generalized profile model parameters for Fornax. }
 \label{fig:result1} 
\end{figure}

\begin{figure}[H]
\centering
\includegraphics[scale=0.55]{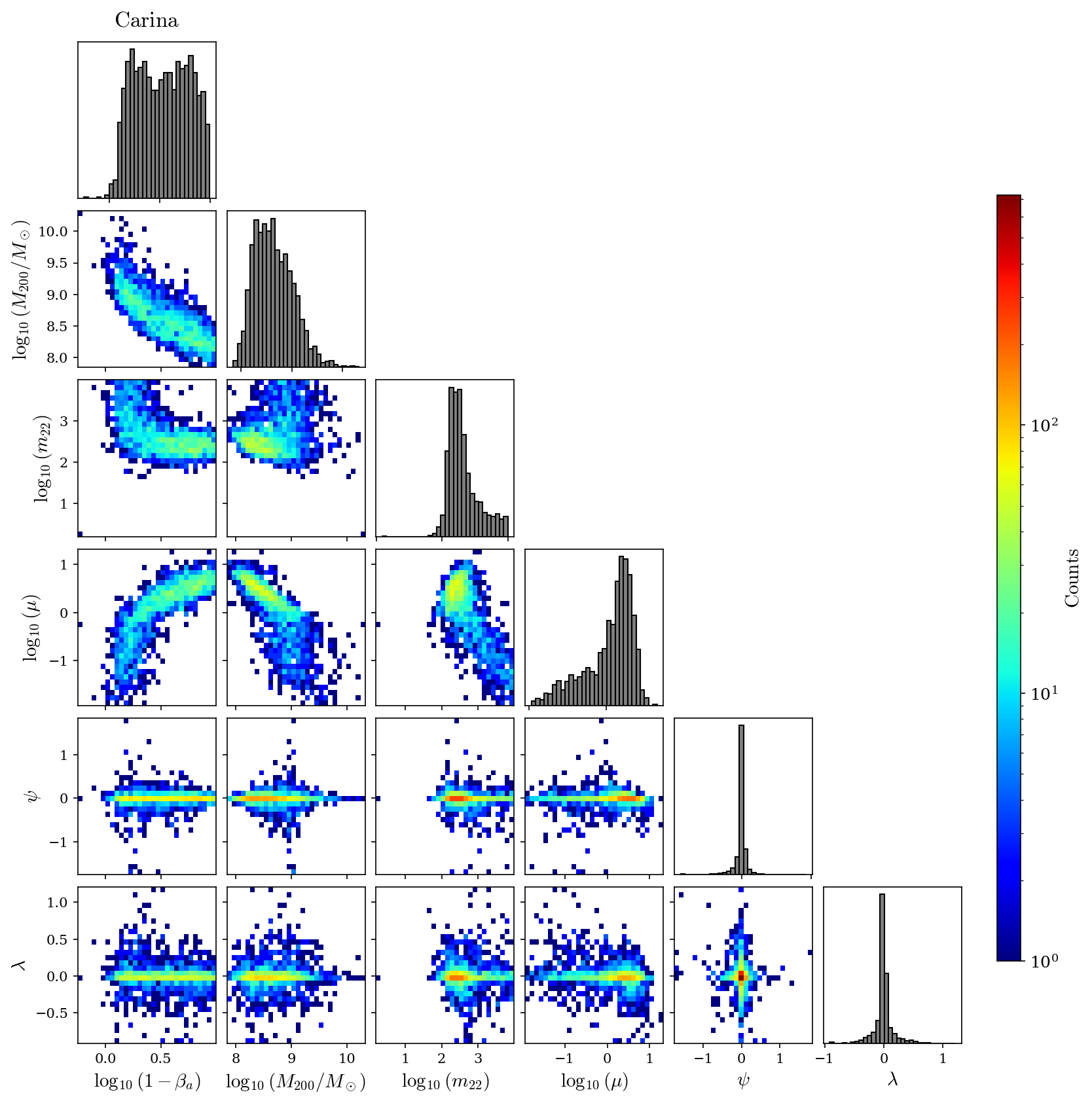}
\caption{Posterior distribution of spin-generalized profile model parameters for Carina. }
 \label{fig:result2} 
\end{figure}

\begin{figure}[H]
\centering
\includegraphics[scale=0.55]{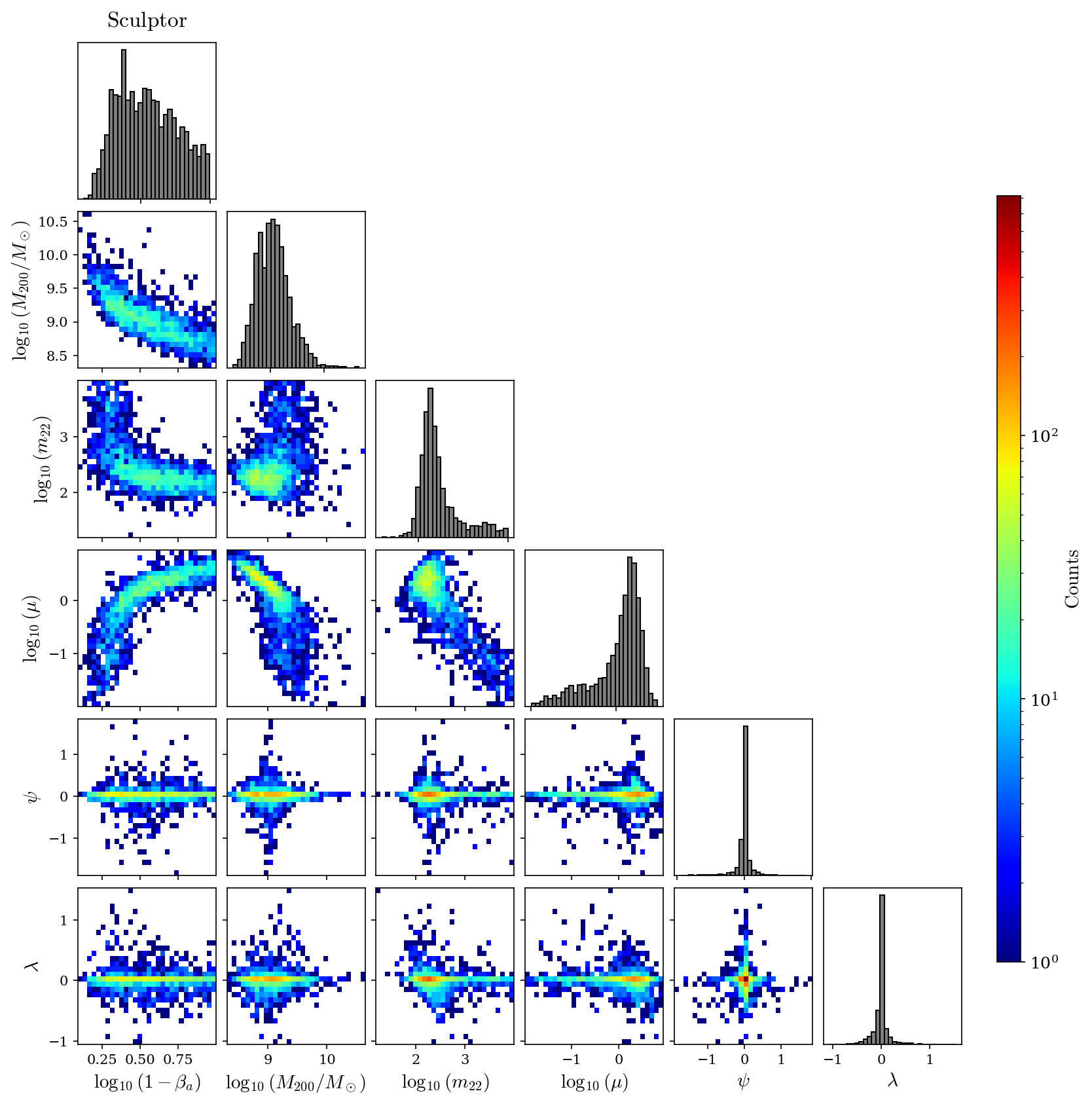}
\caption{Posterior distribution of spin-generalized profile model parameters for Sculptor. }
\label{fig:result3} 
\end{figure}

\begin{figure}[H]
\centering
\includegraphics[scale=0.55]{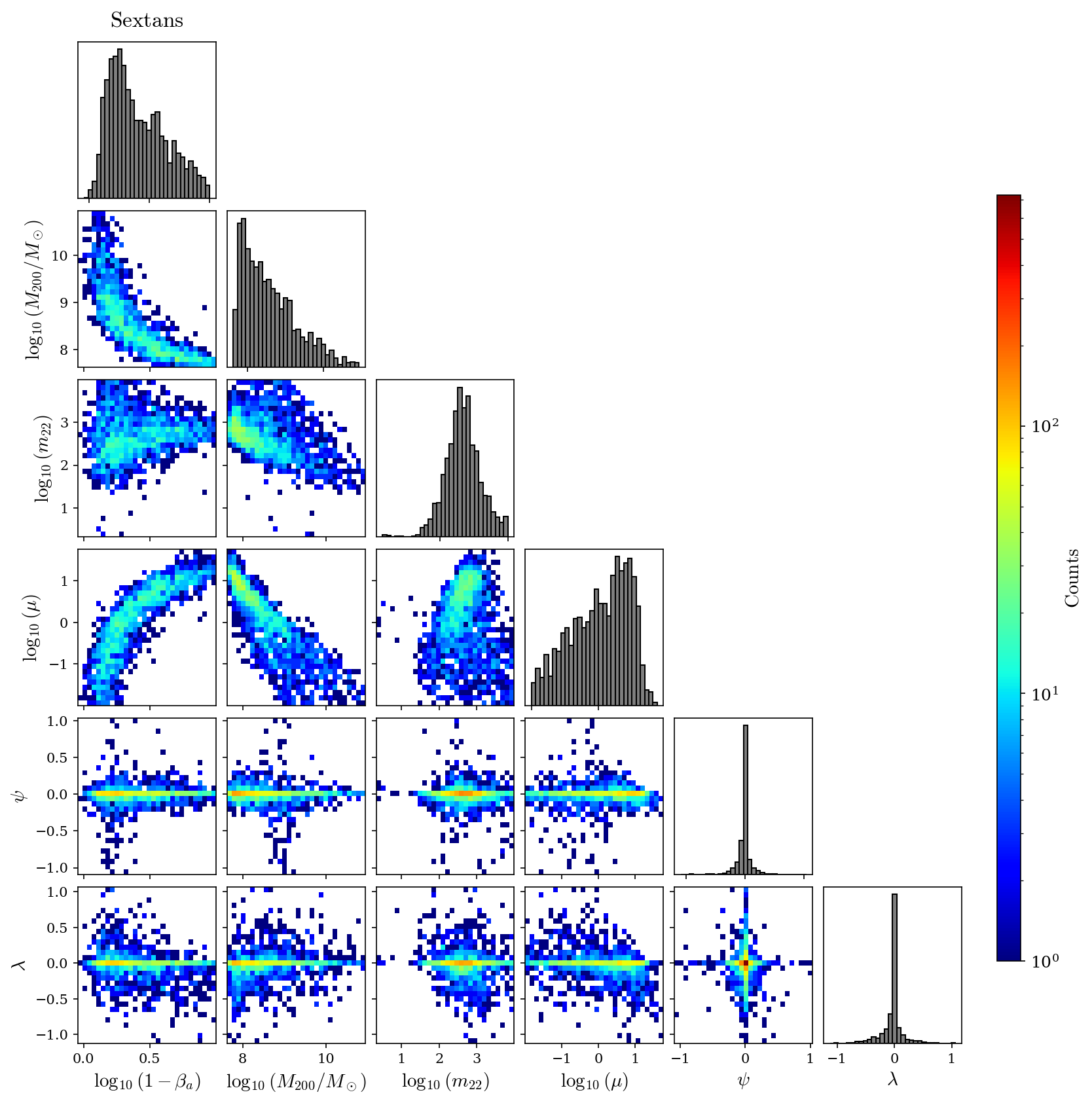}
\caption{Posterior distribution of spin-generalized profile model parameters for Sextans.}
\label{fig:result4} 
\end{figure}

\begin{figure}[H]
\centering
\includegraphics[scale=0.55]{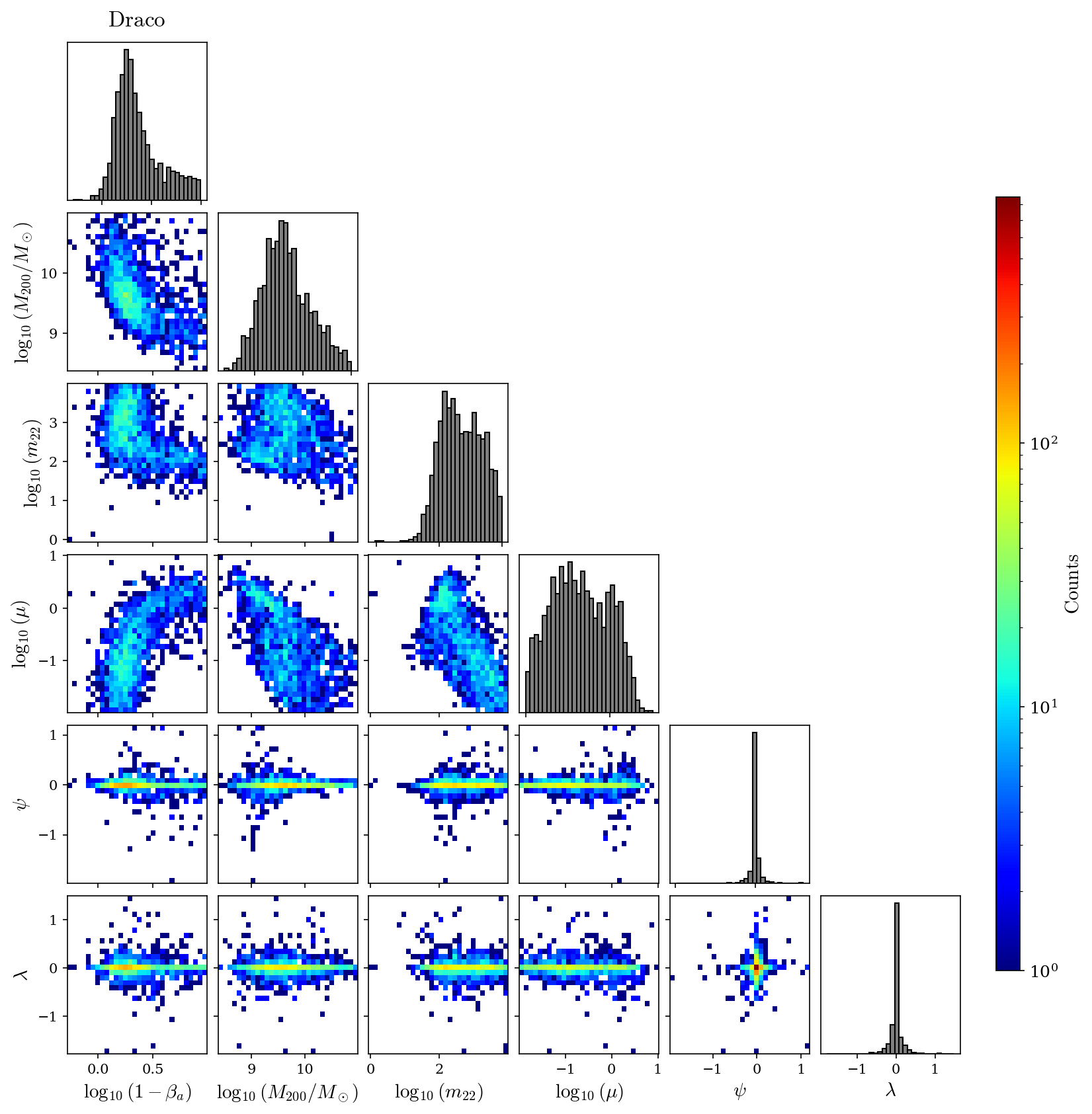}
\caption{Posterior distribution of  spin-generalized profile model parameters for Draco. }
\label{fig:result6} 
\end{figure}

\begin{figure}[H]
\centering
\includegraphics[scale=0.55]{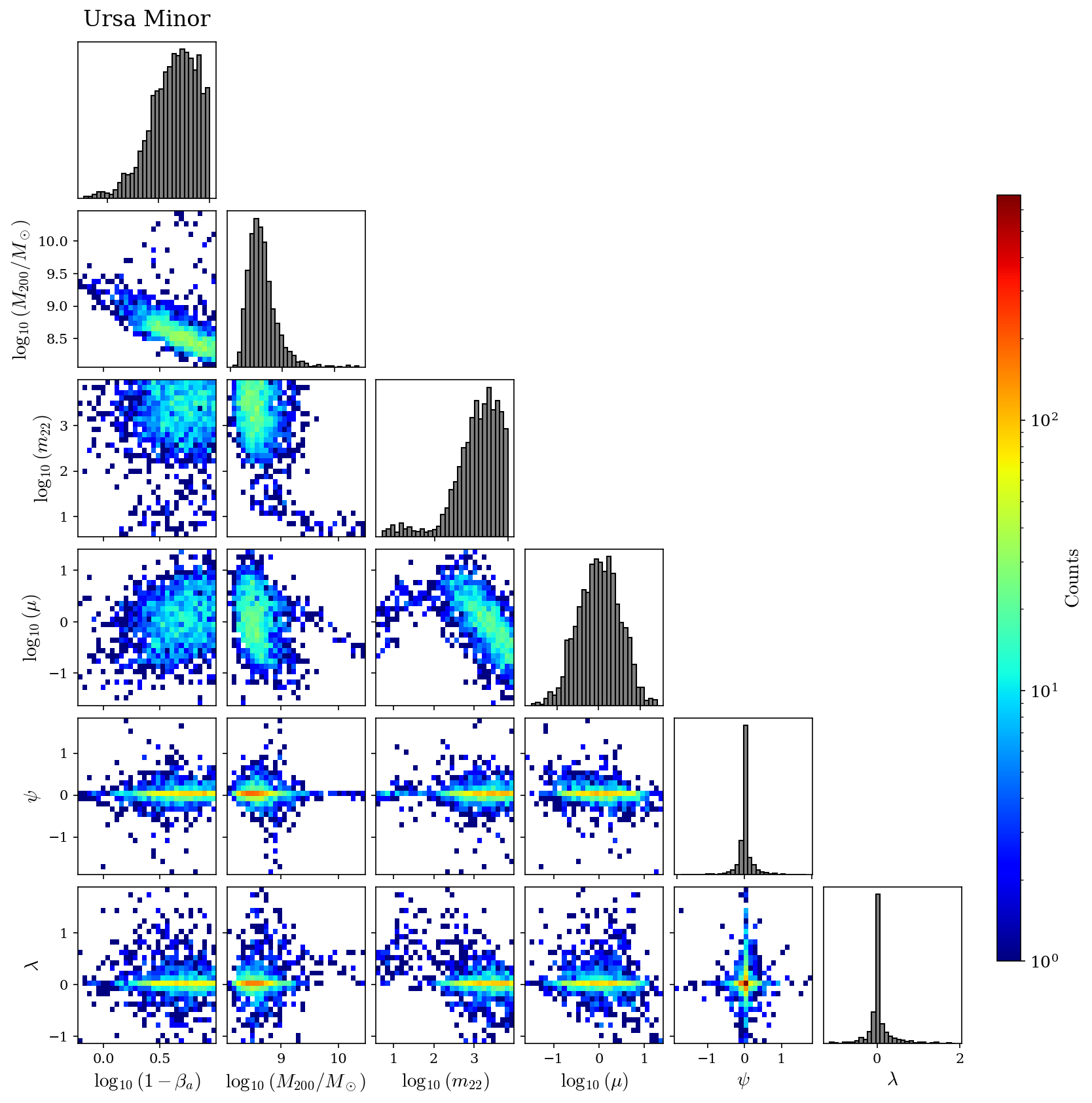}
\caption{Posterior distribution of spin-generalized profile model parameters for Ursa Minor. }
\label{fig:result5} 
\end{figure}

\bibliography{ref}
\end{document}